# Highly Efficient Hyperbranched CNT Surfactants: Influence of Molar Mass and Functionalization

Ellen Bertels,*,[†,‡] Kevin Bruyninckx,[†] Mert Kurttepeli,[§] Mario Smet,[†] Sara Bals,[§] and Bart Goderis[†]

[†]Polymer Chemistry & Materials, KU Leuven, Celestijnenlaan 200F, b2404 Heverlee, Belgium
[‡]Program Nanoforce, SIM Flanders vzw, Technologypark 935, Zwijnaarde, Belgium
[§]EMAT, University of Antwerp, Groenenborgerlaan 171, Antwerp, Belgium

*Supporting Information

**ABSTRACT:** End-group-functionalized hyperbranched polymers were synthesized to act as a carbon nanotube (CNT) surfactant in aqueous solutions. Variation of the percentage of triphenylmethyl (trityl) functionalization and of the molar mass of the hyperbranched polyglycerol (PG) core resulted in the highest measured surfactant efficiency for a 5000 g/mol PG with 5.6% of the available hydroxyl end-groups replaced by trityl functions, as shown by UV−vis measurements. Semiempirical model calculations suggest an even higher efficiency for PG5000 with 2.5% functionalization and maximal molecule specific efficiency in general at low degrees of functionalization. Addition of trityl groups increases the surfactant−nanotube interactions in comparison to unfunctionalized PG because of $\pi-\pi$ stacking interactions. However, at higher functionalization degrees mutual interactions between trityl groups come into play, decreasing the surfactant efficiency, while lack of water solubility becomes an issue at very high functionalization degrees. Low molar mass surfactants are less efficient compared to higher molar mass species most likely because the higher bulkiness of the latter allows for a better CNT separation and stabilization. The most efficient surfactant studied allowed dispersing 2.85 mg of CNT in 20 mL with as little as 1 mg of surfactant. These dispersions, remaining stable for at least 2 months, were mainly composed of individual CNTs as revealed by electron microscopy.

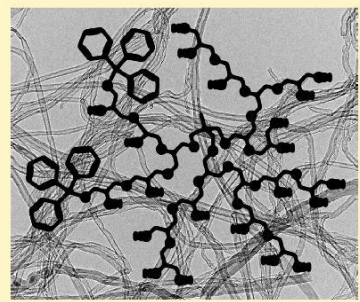

## INTRODUCTION

In 1991 Iijima et al. discovered carbon nanotubes (CNTs) cylindrical structures resembling rolled-up graphite sheets ideally being defect-free with $sp^2$-hybridized carbon−carbon bonds and extended conjugation.[1,2] Because of this structure, these one-dimensional nanofibers possess unique mechanical, optical, electronic, and thermal (conductivity) properties.[2−4] They are used for e.g. biosensors and molecular scale wires because of their electronic (and thermal) properties, whereas their mechanical properties make them interesting candidates as polymer or composite reinforcements. Literature also mentions combinations of both when e.g. used as nanofiller in polymeric composites with improved mechanical performance and possessing electrical conductivity.[5−7] Nonetheless, CNTs tend to entangle and to aggregate due to their structure, which results in the formation of nanoropes and bundles, structures with inferior properties compared to isolated tubes.[4,8]

The main route available to engineering nanotubes for electronic applications as well as nanocomposites and multifunctional devices is solution-based processing,[9−11] which requires tubes to be dispersed in liquids such as organic solvents or surfactant solutions.[8,12,13] Preferentially this results in the formation of a stable dispersion or solution, thereby overcoming the strong intertube $\pi-\pi$ stacking responsible for bundle formation.[14] Nanotube exfoliation requires mechanical methods such as ultrasonication (US) or high shear mixing to break bundles apart, but nonetheless (most frequently) additional substances have to be added to keep the individualized tubes stable[15] Different solvents, often consisting of molecules having $\pi$-electrons, can overcome this issue, e.g., dimethylformamide and N-methylpyrrolidinone but also 1,2-dichlorobenzene, chloroform, and tetrahydrofuran.[6,13,16,17] However, most of the previously mentioned solvents are toxic, and e.g. alcoholic systems, important in sol−gel chemistry and condensation polymerization, can be an alternative, although currently especially aqueous solutions draw attention because of environmental and safety reasons.[3,12,18] When using aqueous solutions, bundles are generally broken apart by ultrasonication, while a physical or chemical treatment renders the individualized tubes or small bundles soluble. Whereas attaching chemical groups changes $sp^2$ into $sp^3$ hybridization, altering properties and physical methods preserve the intrinsic CNT structure and thus e.g. electronic performance.[3,19,20] The latter methods usually consist in the use of surfactants, synthetic polymers, or even natural macromolecules acting as a dispersing agent, diffusing and adsorbing along the tube in spaces or gaps created by US, thereby stabilizing the individual tube.[12,13,21,22] It is believed that longer sonication times and increased sonication strength favor the breaking up of the bundles, exposing additional surface area for the adsorption of surfactant molecules.[15,22]

Dispersing agents consist of a part interacting with the nanotube and a (hydrophilic) group for interaction with the solvent. They counteract attractive potentials between tubes by decreasing $\pi-\pi$ interactions,[3,10,11] while electrostatic or steric repulsion prevents nanotubes from approaching and reagglomerating and reaggregating.[12,17,20,23] Care has to be taken to control the surfactant concentration, as depletion interaction may come into play above the surfactant critical micelle concentration (CMC). Micelles do not fit in between nanotubes, creating lower micelle concentrations at these locations. This leads to a concentration and consequential osmotic pressure difference in the solution, pushing nanotubes together and causing reagglomeration.[24,25] Depletion interaction received quite some attention in research, but disagreement among authors exists as whether to work above[3,14,23] or below[26,27] the CMC of surfactants. In any case, even in the absence of depletion interaction needlessly high surfactant amounts should be avoided for economic reasons and to avoid performance losses when used at a later stage in e.g. composites.

Literature often mentions molecules with basic nitrogen atoms possessing high affinity toward carbon nanotubes,[16,28,29] and also compounds having $\pi$-electrons ($\pi$-aromatic compounds) are suitable to be used as a surfactant.[16,21,30–32] Sodium dodecyl sulfate (SDS) is one of the most popular stabilizing agents for nanotube dispersion in aqueous media[10,13,25] as well as its derivative (sodium) dodecylbenzenesulfonate ((Na)DDBS).[3,33,34] For the latter $\pi-\pi$ stacking interactions between a phenyl group and the CNT surface increase the binding and efficiency of the surfactant.[35,36] Among polymer stabilizers especially poly(ethylene oxide) derivatives such as the block copolymers Brij,[37] Tween,[37] and Pluronic[21] are frequently used as well as natural macromolecules as DNA.[19] Adsorption of those polymeric species also takes place along the nanotube sidewalls, mostly driven by hydrophobic effects and charge transfer interactions with the surfactant lone electron pairs.[19,38]

Zhang already mentioned amphiphilic block copolymers with dendritic structures to be more efficient oil emulsifiers compared to their linear counterparts because of increased penetrability and good absorption and displacement behavior.[39] Xin compared Pluronic, a block copolymer of poly(ethylene oxide) (PEO) and poly(propylene oxide) (PPO), with its five-branched starlike counterpart, having a similar hydrophilic PEO fraction, and found out that the latter had a more pronounced surface activity leading to a higher efficiency, even at lower concentrations of the branched molecule.[38] These branched species improve steric repulsions as confirmed by simulations and have a higher number of hydrophilic groups extending into water per unit of nanotubes.[15] More hyperbranched and dendrimeric surfactants can be found in the literature,[15,40] generally having a hydrophobic core or possessing blocklike arms. The review by Sun et al. lists some examples of hyperbranched polymers of which the main chain contains aromatic moieties, thereby aiming at $\pi-\pi$ stacking inter-actions.[41] To the best of our knowledge, to date no research has considered hyperbranched polyglycerol (PG) with triphenyl-methyl (trityl) end-group functionalization as a surfactant. Such molecules are easily synthesized compared to hyperbranched molecules with aromatic backbones and do not rely on a covalent modification of the CNTs. This paper reports on the influence of the polyglycerol molar mass and of the percentage of trityl end-group functionalization with respect to the surfactant's efficiency to disperse multiwalled carbon nanotubes (MWCNT) in water. The trityl groups are expected to interact with the nanotube surface, while the hyperbranched poly-glycerol mainly should account for water solubility and steric repulsion to prevent individualized nanotubes or small bundles from reapproaching. Trityl groups possess three sets of conjugated $\pi$-electrons that may improve interaction with the CNT surface and hence dispersion efficiency.[42] The efficiency of a single phenyl group was confirmed by different authors by comparing SDS and DDBS,[14,21,34,43] even though the phenyl group of the latter is positioned at the hydrophilic end of the molecule.[23] This entity is also assumed to assist the initial separation of an individual nanotube from the bundle adsorbing laterally on the nanotube surface in the narrow space between adjacent tubes.[23,44] Furthermore, charge transfer interactions between the PG oxygen electron lone pairs and the carbon nanotube surface may take place, as well as hydrophobic effects, analogous to PPO–CNT interactions.[15,45] Meanwhile, the flexibility of this (trityl-functionalized) hyperbranched molecule can enable interaction with the CNT surface without compromising steric repulsion properties.

## EXPERIMENTAL SECTION

Hyperbranched polylglycerol (PG) was obtained from the institute of Organic Chemistry of Johannes Gutenberg University Mainz (Mainz, Germany).[46] Other reagents for synthesis were purchased from Sigma-Aldrich (Steinheim, Germany) unless specified otherwise and were at least of analytical grade. The degree of functionalization of the trityl-functionalized PG was determined by nuclear magnetic resonance ($^1$H NMR) (Bruker Avance 300, Bruker, Billerci, MA) (300 MHz) at room temperature in DMSO-$d_6$ with $\delta$ = 7.5−6.8 (m, −C$_6$H$_5$ from trityl), 4.9-4.3 (m, −OH), and 3.8−3.2 (m, CH and CH$_2$ from the PG) (see Supporting Information). A sample based on PG with a molar mass of y and x% of the available end-groups functionalized is referred to as x% PGy.

Multiwalled carbon nanotubes (NC7050) and acid-functionalized multiwalled carbon nanotubes (NC3151) (both p-type) were purchased from Nanocyl Belgium, and the desired amount a was put in a 20 mL vial, followed by the addition of b surfactant and 20 mL of deionized water, resulting in an a/b ratio. All samples throughout this paper contain 20 mL of water. While sonicating this solution (UP200S US processor, Hielscher Technology) for 60 min with a cycle of 0.65 and an amplitude of 70%, samples were taken and analyzed at regular time intervals. These intervals are referred to as ultrasonication time (US time). The analysis by UV−vis (Lambda 900, PerkinElmer) happened in a 1.00 mm cuvette (QS high precision cell, Helma Analytics) to evaluate solubility and surfactant efficiency. The absorbance at wavelengths from 700 nm down to 200 nm was measured and corrected by subtracting a surfactant solution baseline. To further avoid interference from the surfactants displaying varying absorption between 200 and 400 nm (see Supporting Information), the absorbance at a wavelength of 500 nm was chosen for further analysis (after the mentioned baseline correction). The amount of dispersed MWCNTS is proportional to the absorbance of the solution.[16] An absorbance of 0.244 and 0.238 for respectively the pristine and the acid-functionalized CNTs was found for every mg of nanotubes dispersed (in 20 mL of water) by combining TEM (displaying full dispersion) with UV−vis (see also Results and Discussion). Sigmoidal fits added to the resulting data only serve as a guide to the eye. Scanning electron microscopy (SEM) samples were prepared by cleaning a stainless steel (SST 304, OCAS, Belgium) substrate ultrasonically (Branson 5510) in ethanol, followed by dipping in a 3-aminopropyltriethoxysilane (APS)/water/ethanol solution (2/88/10 v%/v%/v%) for 60 s and rinsing in an ethanol bath. Next this sample was placed in a CNT/surfactant solution for at least 45 min, blown dry with pressurized air and cured in a vacuum oven at 70 °C for 1.5 h. Analysis was done with a FEI Quanta microscope operated at 10 keV. Transmission electron microscopy (TEM) samples were prepared by suspending a CNT/surfactant solution in ethanol, followed by deposition of a drop of this suspension on a carbon-coated TEM grid. TEM and high-resolution TEM (HRTEM) micrographs were collected from a FEI Tecnai F20 operated at 200 kV.

## ■ RESULTS AND DISCUSSION

**Surfactant Synthesis and Characterization.** Synthesis of trityl-functionalized PG started from a PG core, further referred to as PG2000, PG5000, or PG10000 (specifications in Table 1), followed by reaction with triphenylmethyl chloride via a nucleophilic substitution as shown in Figure 1. A mass of 0.2 g of the desired PG was mixed with triphenylmethyl chloride in pyridine (10 mL) and stirred at room temperature for 48 h under an argon atmosphere, followed by slow addition of the polymer mixture to a stirred solution of ethyl acetate (100 mL), resulting in a polymer suspension. Centrifugation of this suspension at 4000g for 5 min yielded a pellet of functionalized PG that was dissolved in methanol (10 mL) and once again added dropwise to ethyl acetate (100 mL). The produced suspension was centrifuged, leading to the final pellet of functionalized PG that was dried in a vacuum oven for a minimum of 12 h at 40 °C. Different degrees of functionalization, i.e., a different percentage of available hydroxyl groups of the starting PG being functionalized with trityl groups, have been synthesized by varying the amount of trityl chloride (see Table 1 for a complete list of fabricated products). Products synthesized in this work starting from PG with molar mass y an with x% of the hydroxyl end-groups functionalized are further referred to as x% PGy. [13]C and [13]C-DEPT-135 NMR spectra provided prove for successful functionalization, while [1]H NMR revealed the degree of functionalization of the trityl-functionalized PG (see Support-ing Information).

Table 1. PG Specifications: Number-Average Molar Mass ($M_n$) of the PG Core, Polydispersity Index (PDI) of the PG Core, Number of Hydroxyl Groups (N) in the PG Core Prior to Functionalization, and the Degree of Trityl End-Group Functionalization (DG) after the Reaction

| name | $M_n$ [g/mol] | PDI | N | DG [%] |
|---|---|---|---|---|
| PG2000 | 1779 | 1.23 | 25 | 0, 2.9, 9 |
| PG5000 | 5456 | 1.57 | 74 | 0, 5.6, 7.8, 13.1, 22.6 |
| PG10000 | 9446 | 1.41 | 126 | 0, 0.6, 2.7, 3.7, 11.2 |

**Nanotube/Surfactant Ratio and MWCNT Functionalization.** UV−vis measurements provide information on the carbon nanotube dispersion as bundled tubes are hardly active in the region measured, whereas individual tubes result in a clear absorption spectrum. The absorbance is linearly related to the amount of individual CNTs as described by the Lambert−Beer law.[14,22] Figure 2 depicts the absorbance at 500 nm during sonication of different masses (a mg) of acid-functionalized multiwalled carbon nanotubes (COOH-MWCNT) for a constant mass (b mg) of surfactant (22.6% PG5000) in 20 mL of water, referred to as a/b 22.6% PG5000. The linear relation between the nanotube concentration and the ultimate absorbance (resulting in an absorbance of 0.24 per mg of nanotubes) suggests maximal dispersion, which is realized irrespective of the COOH-MWCNT/surfactant proportion, at least up to a ratio of 1/1.[12]

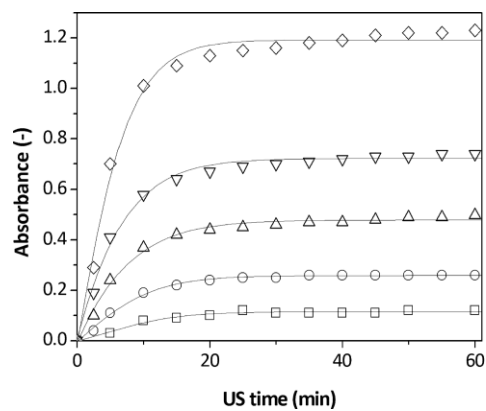

Figure 2. Absorbance of different COOH-MWCNT concentrations as a function of the ultrasonication time (US time) below the critical ratio for a COOH-MWCNT/22.6% PG5000 ratio of 0.5/5 (□), 1/5 (○), 2/5 (△), 3/5 (▽), and 5/5 (◇).

Absence of surfactant leads to zero absorbance as no individual tubes can be brought into dispersion. At the other extreme of excessively high surfactant concentrations and low surfactant/MWCNT ratios depletion aggregation, as described above, may interfere and reduce the solubilizing capacity. However, no micelles were found by DLS for 5.6% PG5000 in water systems for concentrations up to 5 g/L. This concentration by far exceeds the CMC of e.g. Brij and Triton,[47] which might be due to the inability of the synthesized molecules to form micelles. Precipitation due to solubility issues may take place rather than the formation of micelles (see further below). Given that the nanotubes consume surfactant during their suspension and thus leave less free surfactant in solution for potential micelle formation, it is conceivable that CNT loaded systems do not contain any micelles.[15] Depletion interactions in the present cases therefore can be excluded.

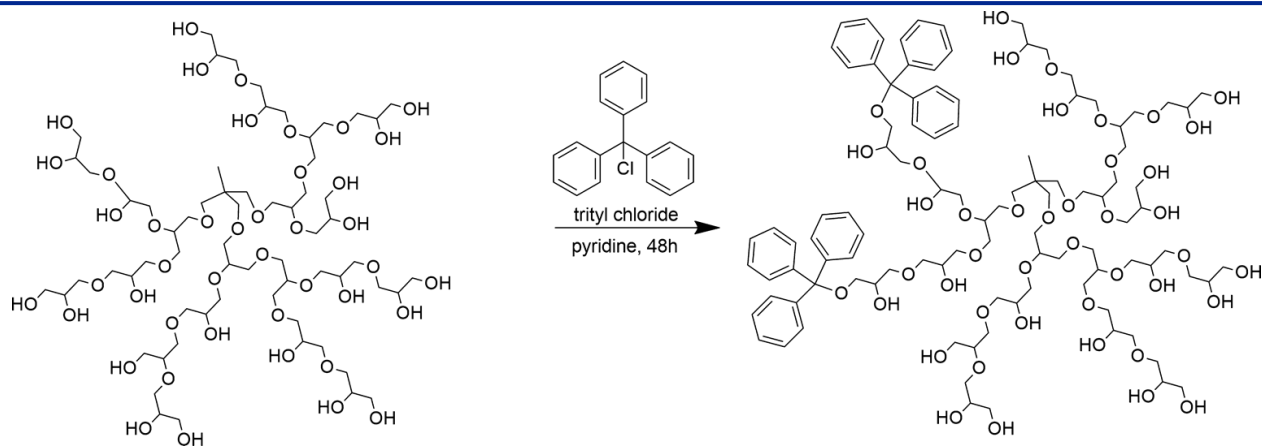

Figure 1. Synthesis of the trityl end-group functionalized hyperbranched polyglycerol.

Studies on MWCNT/surfactant ratios above a certain value (see Figure 3) demonstrate that a critical ratio exists above which insufficient surfactant molecules are present to stabilize all individual tubes peeled off during ultrasonication.[48] This holds also for acid-functionalized MWCNTs. At first the relevant experiments related to nonfunctionalized MWCNTs will be discussed. Systems with COOH-MWCNTs will be compared to those with nonfunctionalized ones at the end of this section.

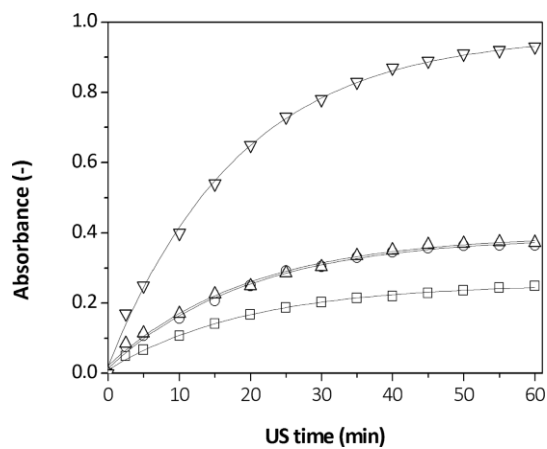

Figure 3. Absorbance around and above the critical ratio as a function of ultrasonication time (US time) for a MWCNT/22.6% PG5000 ratio of 1/1 (□), 2/1 (○), 3/1 (△), and 5/2.5 (▽).

Figure 3 shows for MWCNT/22.6% PG5000 systems that the 2/1 ratio has an absorbance slightly lower than twice the value of its 1/1 counterpart and that the absorbance of a 3/1 sample levels off around the same value as the 2/1 ratio does. The 1/1 proportion seems to be below the critical ratio for this surfactant given the absorbance of 0.24 per mg of nanotubes added. Moreover, this solution remains stable for at least 2 months (data not shown). The 2/1 and 3/1 ratios exceed the critical ratio as deduced from their absorbance being lower than 0.24 per mg of MWCNT added. Although both the 2/1 and 5/2.5 ratio slightly exceed the critical ratio and thus contain some remaining bundles, their comparison demonstrates that higher surfactant concentrations can disperse more nanotubes. The absorbance of the latter is exactly 2.5 times that of the 2/1 ratio, confirming the existence of a critical ratio rather than a critical absolute surfactant concentration in the studied concentration range. Because the maximum dispersed nanotube concentration depends on the available surfactant, knowledge of the critical ratio is of major importance as higher dispersed nanotube concentrations can only be realized by increasing the surfactant concentration such that the ratio tumbles below the critical one.[34]

Most authors agree that higher surfactant concentrations result in the potential of dispersing larger amounts of nanotubes.[14,15,24] Some on the contrary mention that the nanotube/surfactant ratio is of minor importance compared to the absolute surfactant concentration.[23,49] Blanch et al. suggested that the increase of CNT dispersion with surfactant concentration is caused by an associated increase of the viscosity.[23] In the present work, however, the absolute surfactant concentration and the associated viscosity effect seem to be irrelevant and the nanotube/surfactant ratio to be dominant.

Figures 2 and 3 demonstrate the absorbance of 0.24 per mg (COOH-)MWCNT added in the case of maximum dispersion. This value can already be found for a ratio of at least 1/1 for the entire PG5000 series, irrespective of the degree of functionalization. Also the other molar mass series demonstrate maximum dispersion for this ratio, with dispersions remaining stable for at least 2 months, except for the 11.2% PG10000 sample (data not shown), rendering these species extremely efficient compared to other surfactants. Literature generally reports values from 1/5 to 1/10 by weight,[14,43] and extreme ratios from 1/2[22,26] down to 1/350[50] occur more exceptionally. Hyperbranched dispersants with PPO−PPE copolymer arms require a CNT/surfactant ratio of 1/3 to disperse nanotubes, whereas a ratio of at most 1/30 is needed to keep these dispersions stable.[15] For Arabic gum, a highly branched polysaccharide, ratios as low a 1/25 have been reported.[40]

Except for Bystrzejewski et al., who claims that the acid oxidation of nanotubes does not influence the dispersed concentration,[14] literature agrees that chemical CNT function-alization results in higher populations of individual tubes,[3] even at very low levels of covalent functionalization (1−2 at. %).[10] Chemical treatment disrupts the CNT structure, decreases π−π-stacking, and increases interactions with the solvent, consequently enabling dispersion. This higher efficiency after CNT functionalization is confirmed in Figure 4, demonstrating that a ratio at least as high as 3/1 leads to most efficient dispersion for COOH-MWCNT when using the PG2000 and PG5000 series. Whereas 11.2% PG10000 clearly disperses less nanotubes (as seen by the lower absorbance), 1 mg of the most efficient surfactant (5.6% PG5000, see also next subsection) can disperse more than 5 mg of COOH-MWCNTs. It must be noted that the highest possible absolute (COOH-)MWCNT loading was not studied, as this leads to absorbances too high to be measured. The next section, discussing trends in the critical ratio, therefore focuses on MWCNT systems, as MWCNT dispersions display saturations in dispersing capabilities already at lower CNT/surfactant ratios compared to COOH-MWCNT systems. Moreover, the former ones are more interesting from an application point of view as they display better mechanical and electrical properties.

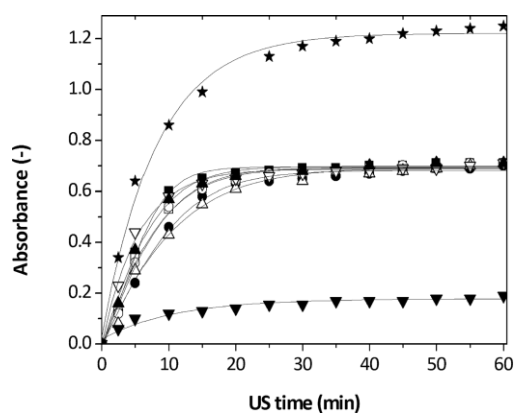

Figure 4. Absorbance of COOH-MWCNT dispersions as a function of ultrasonication time (US time) for 3/1 MWCNT/2.9% PG2000 (■), 3/1 MWCNT/9% PG2000 (●), 3/1 MWCNT/5.6% PG5000 (□), 3/1 MWCNT/7.88% PG5000 (○), 3/1 MWCNT/13.1% PG5000 (△), 3/1 MWCNT/22.6% PG5000 (▽), 3/1 MWCNT/ 0.6% PG10000 (⊞), 3/1 MWCNT/2.7% PG10000 (⊗), 3/1 MWCNT/3.7% PG10000 (▲), 3/1 MWCNT/11.2% PG10000 (▼), and 5/1 MWCNT/5.6% PG5000 (★).

Influence of PG Molar Mass and End-Group Function-alization on Surfactant Efficiency. Figure 5 depicts the maximum MWCNT mass (mg) dispersed per 1 mg of surfactant in 20 mL of water, a quantity further referred to as the surfactant efficiency. These efficiencies are numerically equivalent to the critical ratio and hence dimensionless (mass MWCNT/mass surfactant). Calculation of the amount of dispersed MWCNT based on the final absorbance after 60 min of sonication, assumed again that an absorbance increment of 0.24 corresponds to 1 mg of dispersed MWCNT. The efficiencies are grouped according to the core PG molar mass and represented as a function of the hydroxyl group percentage replaced with trityl functionalities. Note here that the 22.6% PG5000, the 3.7%, and 11.2% PG10000 are not fitted nor used in further calculations as aqueous solutions of this surfactant appear hazy, pointing at solubility issues thereby potentially changing the active surfactant concentration. Figure 5 also includes the three unfunctionalized PG species, which clearly are already fairly efficient when compared to e.g. SDS (also included), confirming interactions between the nanotubes and the PG core. As mentioned in the Introduction, this can be due to charge transfer interactions between the PG oxygen electron lone pairs and the carbon nanotube surface and perhaps also partially to small hydrophobic effects.[15,45] PG2000 has a lower efficiency compared to PG5000 and PG10000, probably because PG2000 is too small to cause a steric stabilization as large as that of its larger counterparts. An increased steric stabilization seems to outweigh the fact that a lower effective number of molecules per unit of mass is introduced when larger molar masses are used, which in principle may lead to a poorer nanotube stabilization.[21]

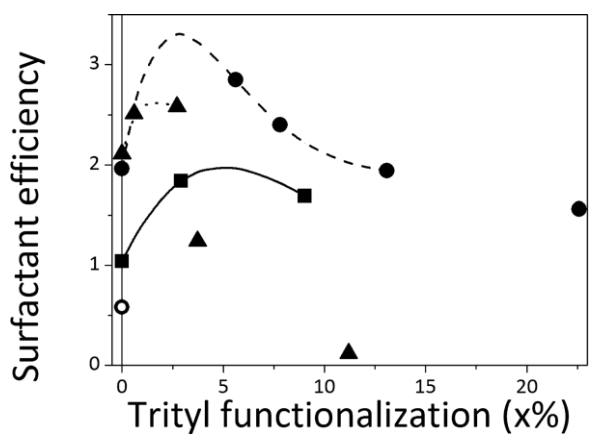

Figure 5. Surfactant efficiency as a function of PG trityl end-group functionalization (x%) for PG2000 (■), PG5000 (●), and PG10000 (▲). Trend lines are fits through the selected data points, leading to the molecular efficiencies displayed in the figure. The data points not being connected by the lines are excluded from the analysis because of surfactant solubility problems.

The present surfactants differ from most others as they are complex mixtures of molecules with different molar masses and degrees of trityl functionalization per molecule. The introduced interpretation scheme attempts to grasp some of these aspects, based on the assumption that the global surfactant efficiency is the average of the molecule specific efficiencies. Furthermore, the treatment focuses on the distribution in trityl functionaliza-tion and neglects the molar mass polydispersity as well as topological differences between molecules with an identical amount of trityl functionalizations. This assumes e.g. all molecules in PG2000 to be equally large, carrying 25 hydroxyl groups available for functionalization (see also Table 1). In addition, molecules with e.g. two functionalizations are considered equally efficient, irrespective of the attachment of the functionalizations: at the periphery, the core of the molecule or in each others proximity or on opposite sides of the PG molecule. Alternatively, a single average efficiency represents all molecules with an identical number of functionalizations. Assuming that the probability of trityl functionalization does not depend on earlier trityl attachments or on whether a primary or secondary alcohol is being replaced, one can compute the probability ($p_i$) that a molecule will carry $i$ functionalizations as follows:

$$p_i = \phi^i (1-\phi)^{N-i} \frac{N!}{(N-i)!\,i!} \quad (1)$$

with $\phi$ the average fraction of functionalizations and $N$ the number of hydroxyl groups available for functionalization in each molecule. With the assumptions listed above, $\phi$ and $N$ for 5% PG2000 equal 0.05 and 25, respectively.

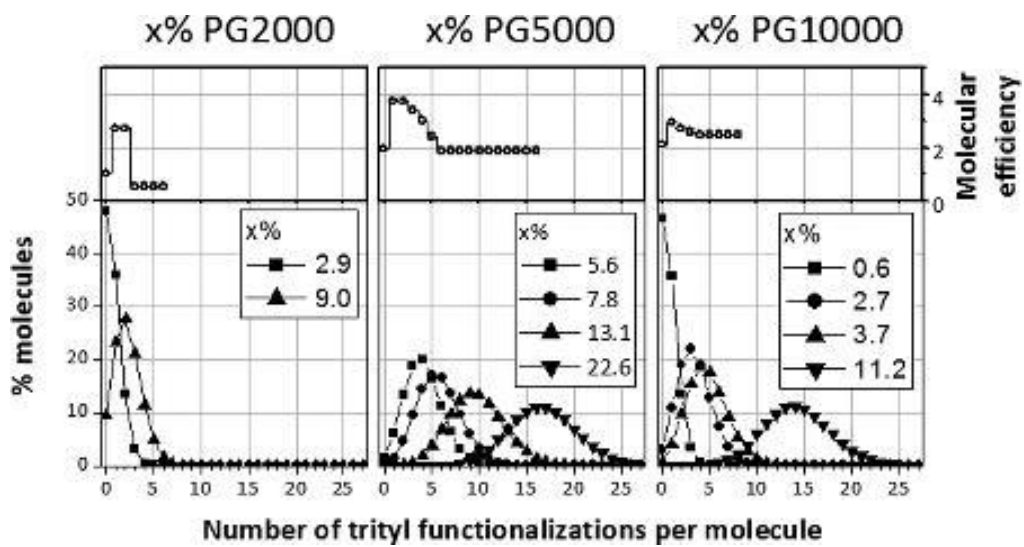

Figure 6. Trityl functionalization distribution for an average functionalization x% (with x% = φ), as indicated in the legends (bottom panels). The molecular efficiency, i.e., the efficiency to disperse MWCNTs for molecules with a given number of trityl functionalizations and PG core (top panels). x% PG2000 species are depicted on the left, x% PG5000 in the middle, and x% PG10000 on the right-hand side.

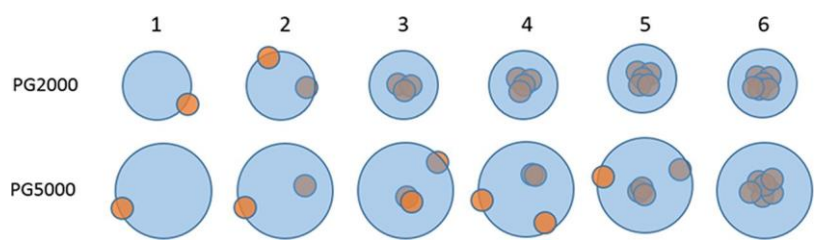

Figure 7. Cartoons representing the PG core (with a size proportional to the PG radius of gyration in water)[51] (blue) and trityl functionalizations (orange) for PG2000 (top row) and PG5000 (bottom row) based species. Beyond a threshold of 2−3 functionalizations the trityl groups tend to mutually interact and do so completely on the average at 3 and 6 functionalizations for respectively PG2000 and PG5000. Mutually interacting trityl groups are suggested to be shielded from the aqueous environment by the PG moieties.

Figure 6 illustrates the result of such calculations for all samples studied, with $p_i$ (expressed as % molecules, i.e., $100p_i$) in function of $i$, the number of trityl functionalizations per molecule. Very clearly, at a low (global) degree of functionalization (x%) and low PG molar mass, a significant percentage of the molecules remain unfunctionalized (up to about 50% in 2.9% PG2000). This unfunctionalized share vanishes at higher PG molar masses and larger x% values. Moreover, the trityl functionalization distributions for a given x% broaden with increasing PG molar mass.

Each of the molecules within the distributions is expected to display its specific molecular efficiency to disperse MWCNTs (top panel Figure 6). These molecular efficiencies were obtained by fitting the experimental global surfactant efficiencies depicted in Figure 5 to the $p_i$ weighted sum of the molecular efficiencies using a least-squares error minimization procedure varying the molecular efficiencies. The minimization procedure was executed combining all x% species of a given PG molar mass in a single run (but excluding the surfactants that display solubility problems, as mentioned before). Given the lower number of experimental data points and to avoid spurious oscillations, the efficiency of molecules with $i$ functionalizations was restricted to be smaller than or equal to the efficiency of molecules with $i-1$ functionalizations, except for molecules with $i = 1$, having no constraint. All calculations were done using the Microsoft Excel solver. The molecular efficiencies for which $p_i$ for all x% is less than 1% are not included in Figure 6 as these partitions do not significantly contribute. Interpolations based on the earlier obtained molecular efficiencies and the x% specific functionalization distribution (eq 1) lead to the trend lines linking the symbols in Figure 5. Clearly, this procedure allows describing the experimental data rather well, and with the aid of Figure 6, aspects related to the PG molar mass and degree of trityl functionalization can be discussed.

The molecular efficiencies for x% PG2000, x% PG5000, and x% PG10000 display a similar trend: at low degrees of functionalization, the efficiency is larger compared to unfunctionalized PG. Beyond a given number of trityl functionalizations, however, the molecular efficiency tumbles down to another (fairly constant) value which for the PG2000 and PG5000 series occurs below that of the parent unfunctionalized PG. This fallback saturates at a trityl functionalization of 3, 6, and 4 for respectively the PG2000, PG5000, and PG10000 series (Figure 6, top panels). The higher molecular efficiency for low trityl functionalizations seems natural and to be related to the anticipated π−π stacking interactions between the trityl phenyl groups and the CNT surface, often found in the literature and confirmed using $^1$H NMR (see Supporting Information).[30,41] As the functionalized species have a higher stability in time (data not shown), trityl−CNT interactions can be considered superior to hydrophobic effects and charge transfer interactions between the surfactant lone pairs and carbon nanotubes. However, the phenyl rings may also interact mutually (see Supporting Information for NMR-based evidence), both inter- and intramolecularly. The probability for intramolecular, mutual interactions increases with decreasing positioning distance between trityl groups, which in turn depends on the number of functionalizations and the size (molar mass) of the parent PG molecule. This effect, starting to take place at 2−3 functionalizations and becoming more probable with increasing degree of functionalization, is illustrated in Figure 7. This figure furthermore suggests that mutually interacting trityl groups may represent a rather hydrophobic entity with a tendency to be shielded from the aqueous environment by the PG moieties. With this shielding effect, no big differences between the efficiencies for functionalizations higher than the mentioned thresholds are expected, as observed in Figure 6. It is conceivable that part of the charge transfer interactions between the PG oxygen electron lone pairs are taking place with the trityl groups, rather than with the MWCNTs, and that this effect is adding to the reduction of the surfactant efficiency below that of pure PG.

Concomitantly to increasing the number of trityl functionalizations, the hydrophilic−lipophilic balance lowers, influencing the surfactant solubility. For the PG10000 series, haziness already starts to set in at a functionalization degree of 3.7%; for the PG5000 series the 22.6% species suffer from solubility issues. The higher efficiency for lower functionalizations thus agrees with what has been found for linear surfactants, i.e., a higher HLB and water solubility increase the surfactant efficiency in aqueous solutions.[21]

The lower molar mass for PG2000 may account for the saturated fallback at 3 compared to at 6 trityl functionalizations for the larger PG5000. It can be verified that the efficiency fallback for these two systems saturates once the total mass corresponding to the trityl functionalizations exceeds about one-third of the PG molar mass. This finding conflicts with papers on nonionic (linear) surfactants stating that equally large hydrophobic and hydrophilic moieties should perform better as a surfactant.[42]

Following the above-mentioned rule of thumb, one expects the efficiency fallback for the PG10000 series to saturate at 16 functionalizations. In contrast, according to Figure 6 this already takes place at four functionalizations. The analysis of the PG10000 series should, however, be considered with some care. Figure 5 demonstrates that the 3.7% and 11.2% PG10000 perform very badly. This can obviously be explained by the low water solubility of the higher functionalized PG10000 species, by which the effective concentration in solution is decreased and thus also the surfactant efficiency. This effect inflates with increasing degrees of functionalization. The 3.7% PG10000 sample in water was clearly hazy and was therefore excluded from the analysis. Although the 0.6 and 2.7% PG10000 samples did not show visible solubility issues, it cannot be excluded that a fraction of the latter (less likely for the 0.6% sample) cannot be dissolved either. If so, the entire analysis for the PG10000 series might be inaccurate. Finally, the model curves in Figure 5 predict a maximal surfactant efficiency for PG5000 with an average functionalization degree of about 2.5%. Therefore, this material deserves consideration in the future.

Most literature mentions a higher efficiency with an increase in molar mass for nonionic surfactants such as the Pluronic, Brij, Tween, and Triton series, due to an increase in steric stabilization and consequently in repulsive forces between individualized tubes.[37,52,53] Vaisman et al. and Xin et al. state that shorter (Pluronic) systems are inefficient in nanotube dispersion not only due to lower molar mass and poor steric hindrance but also due to shorter terminal hydrophilic groups and lower solubility.[15,21] Nonetheless, Vaisman et al. added that enlarging the surfactant hydrophilic part decreases adsorption ratios onto the nanotubes.[21,42] The mentioned trends are partially confirmed in Figure 5 where increasing the polyglycerol molar mass at lower degrees of functionalization from 2000 up to 5000 g/mol improves the nanotube dispersion efficiency. At a molar mass of 10 000 g/mol the trend inverts, likely because of the surfactant solubility issues.

**Structural Aspects.** Figure 8 displays different systems with a 2/1 MWCNT/surfactant ratio using respectively 5.6% PG5000 (A), 2.9% PG2000 (B), and 3.7% PG10000 (C). Only in Figure 8A the MWCNT/surfactant ratio lays below the critical one. Consequently, the system appears very homogeneous and suggests a complete dispersion. In contrast, the system in Figure 8B contains some fine dust and that in Figure 8C clearly suffers from strong flocculation.

The relative absorbance of 0.24 per of nanotubes added for a MWCNT/5.6% PG5000 system at its critical ratio (i.e., 2.85/1) remained so far at least 2 months (data not shown), pointing at the stability of these dispersions. A more detailed investigation using electron microscopy confirms the existence of individual, exfoliated nanotubes in these systems and hence the surfactant efficiency. SEM images of nanotubes deposited from PG5000 solutions on APS-covered stainless steel substrates (Figure 9A) clearly reveal dispersed and individual nanotubes. In a next step these MWCNT covered substrates can be used for manufactur-ing e.g. polymer−steel hybrids with improved mechanical properties.[54] Similar individualized MWCNTs can be found in the TEM image of Figure 9B, collected from a CNT/5.6% PG5000 dispersion at its critical ratio. Figure 9C (HRTEM) shows an individual tube with an interplanar spacing of 0.34 nm, the typical graphene interlayer distance in graphite.

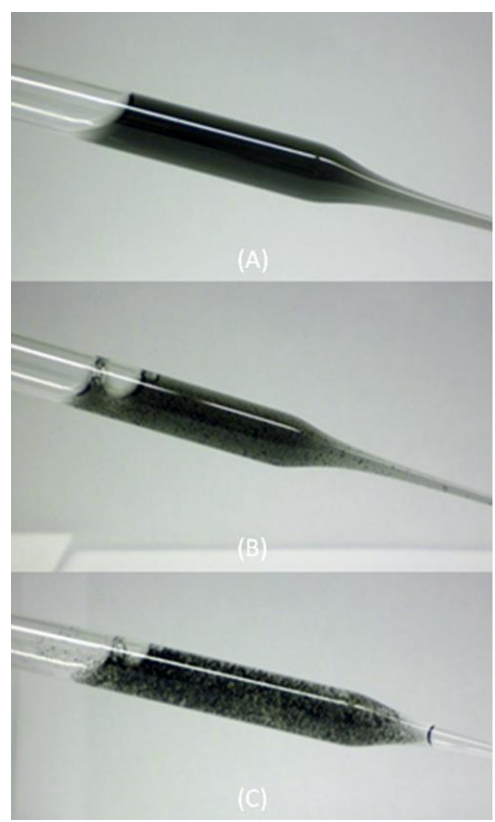

**Figure 8.** Visual inspection of dispersions with (A) 5.6% PG5000, (B) 2.9% PG2000, and (C) 3.7% PG10000. For all samples 2 and 1 mg of respectively MWCNT and the surfactant were brought into 20 mL of water and sonicated for 60 min.

## ■ CONCLUSION

Hyperbranched trityl-functionalized polyglycerol surfactants can be synthesized as desired, offering flexibility in the percentage of end-group functionalization as well as in polyglycerol molar mass. The surfactant efficiency expressed in terms of the critical CNT/surfactant ratio for these species clearly depends on the composition of the surfactant. UV−vis absorbance measurements show high dispersion power, even for the unfunctionalized species, confirming charge transfer interactions with the surfactant lone pairs. Addition of trityl groups increases efficiency and long-term stability due to $\pi-\pi$ stacking interactions, the existence of which is confirmed by NMR. Semiempirical model calculations point at maximal molecule specific efficiency at low degrees of functionalization due to an increased surfactant−nanotube interaction compared to unfunctionalized PG. At too high functionalization degrees, mutual interactions between trityl groups come into play (seen by NMR), decreasing the surfactant efficiency, while lack of water solubility becomes an issue for higher molar mass species at higher degrees of functionalization. Low molar mass surfactants are less efficient compared to higher molar mass species most likely because the higher bulkiness of the latter allows for a better CNT separation and stabilization.

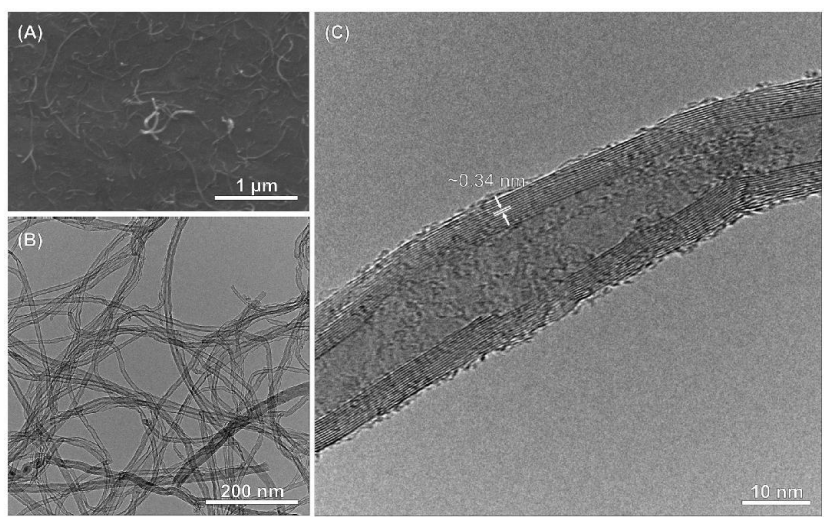

Figure 9. Top-view SEM micrograph showing MWCNT deposited from solution on an APS-coated stainless steel substrate (A). TEM image collected from a CNT/5.6% PG5000 solution (B). MWCNTs as visualized by HRTEM (C).

The most efficient surfactant studied (5.6% PG5000) allowed dispersing 2.85 mg of CNT in 20 mL of water with as little as 1 mg of surfactant, although calculations indicate an even higher dispersion power for PG5000 with a slightly lower degree of functionalization. Electron microscopy revealed the dispersions, remaining stable for at least 2 months, were mainly composed of individual CNTs.

In the future computational simulations of surfactant behavior in water as a function of the degree of functionalization may provide extra support for the suggested intramolecular trityl group interactions and may explain the gradual fallback in surfactant efficiency and help in the design of even more efficient analogues. From an application point of view these surfactants may enable the dispersion of CNT in polymeric matrices, thereby improving the mechanical properties of CNT/polymer composites. Here the hydrophobic trityl moieties interact with the tube, while the hyperbranched part can interact with the (polar) polymer through hydrogen bonding or covalent reaction, making it act as a coupling agent.[34,40,55] However, optimizing the surfactant as a coupling agent may still require adaptations to the polyglycerol core.

## ASSOCIATED CONTENT

*Supporting Information

$^1$H and $^{13}$C NMR spectra for unfunctionalized PG10000 as well as 3.7% PG10000; $^{13}$C NMR spectrum with peak designations for 3.7% PG10000 (in DMSO); aromatic part of the $^1$H NMR spectrum for different species synthesized in DMSO (demon-strating trityl group interactions) and aromatic part of this spectrum for a 2.7% PG10000 aqueous solution before and after the addition of nanotubes and ultrasonication (demon-strating surfactant−CNT interactions); UV−vis spectra of an aqueous surfactant solution and this solution after addition of nanotubes and ultrasonication. This material is available free of charge via the Internet at http://pubs.acs.org.

## AUTHOR INFORMATION

Corresponding Author
*E-mail: ellen.bertels@chem.kuleuven.be (E.B.).

Notes
The authors declare no competing financial interest.

## ACKNOWLEDGMENTS

The authors gratefully acknowledge the SIM NanoForce programme for their financial support and thank the group of Prof. Thierry Verbiest, especially Maarten Bloemen, for the use of their UV−vis equipment. Bart Goderis and Mario Smet thank KU Leuven for financial support through a GOA project. Mert Kurttepeli and Sara Bals acknowledge funding from the European Research Council under the 7th Framework Program (FP7), ERC Starting Grant No. 335078 COLOURATOMS.

## REFERENCES

(1) Iijima, S. Helical microtubules of graphitic carbon. Nature 1991, 354, 56−58.
(2) Chen, H.; Jacobs, O.; Wu, W.; Ruediger, G.; Schaedel, B. Effect of dispersion method on tribological properties of carbon nanotube reinforced epoxy resin composites. Polym. Test. 2007, 26, 351−360.
(3) Coleman, J. N. Liquid-phase exfoliation of nanotubes and graphene. Adv. Funct. Mater. 2009, 19, 3680−3695.
(4) Gojny, F.; Wichmann, M.; Fiedler, B.; Schulte, K. Influence of different carbon nanotubes on the mechanical properties of epoxy matrix composites - A comparative study. Compos. Sci. Technol. 2005, 65, 2300−2313.
(5) Gojny, F.; Wichmann, M.; Fiedler, B.; Kinloch, I.; Bauhofer, W.; Windle, A.; Schulte, K. Evaluation and identification of electrical and thermal conduction mechanisms in carbon nanotube/epoxy compo-sites. Polymer 2006, 47, 2036−2045.
(6) Liu, T.; Phang, I.; Shen, L.; Chow, S.; Zhang, W. Morphology and mechanical properties of multiwalled carbon nanotubes reinforced nylon-6 composites. Macromolecules 2004, 37, 7214−7222.
(7) White, C.; Todorov, T. Carbon nanotubes as long ballistic conductors. Nature 1998, 393, 240−242.
(8) Coleman, J. N.; Khan, U.; Blau, W. J.; Gun'ko, Y. K. Small but strong: A review of the mechanical properties of carbon nanotube-polymer composites. Carbon 2006, 44, 1624−1652.
(9) Safadi, B.; Andrews, R.; Grulke, E. Multiwalled carbon nanotube polymer composites: Synthesis and characterization of thin films. J. Appl. Polym. Sci. 2002, 84, 2660−2669.

(10) Usrey, M. L.; Strano, M. S. Adsorption of single walled carbon nanotubes onto silicon oxide surface gradients of 3-aminopropyltri-(ethoxysilane) described by polymer adsorption theory. Langmuir 2009, 25, 9922−9930.

(11) Liu, J.; Casavant, M.; Cox, M.; Walters, D.; Boul, P.; Lu, W.; Rimberg, A.; Smith, K.; Colbert, D.; Smalley, R. Controlled deposition of individual single-walled carbon nanotubes on chemically function-alized templates. Chem. Phys. Lett. 1999, 303, 125−129.

(12) Rouse, J. Polymer-assisted dispersion of single-walled carbon nanotubes in alcohols and applicability toward carbon nanotube/sol-gel composite formation. Langmuir 2005, 21, 1055−1061.

(13) Sa, V.; Kornev, K. G. Analysis of stability of nanotube dispersions using surface tension isotherms. Langmuir 2011, 27, 13451−13460.

(14) Bystrzejewski, M.; Huczko, A.; Lange, H.; Gemming, T.; Buechner, B.; Ruemmeli, M. H. Dispersion and diameter separation of multi-wall carbon nanotubes in aqueous solutions. J. Colloid Interface Sci. 2010, 345, 138−142.

(15) Xin, X.; Xu, G.; Zhao, T.; Zhu, Y.; Shi, X.; Gong, H.; Zhang, Z. Dispersing carbon nanotubes in aqueous solutions by a starlike block copolymer. J. Phys. Chem. C 2008, 112, 16377−16384.

(16) Bahr, J.; Mickelson, E.; Bronikowski, M.; Smalley, R.; Tour, J. Dissolution of small diameter single-wall carbon nanotubes in organic solvents? Chem. Commun. 2001, 193−194.

(17) Chao-Xuan, Liu; Jin-Woo, Choi Improved dispersion of carbon nanotubes in polymers at high concentrations. Nanomaterials 2012, 2, 329−47.

(18) Brinker, C.; Schere, G. Sol-Gel Science: The Physics and Chemistry of Sol-Gel Processing; Academic Press: New York, 1990.

(19) Kim, S. W.; Kim, T.; Kim, Y. S.; Choi, H. S.; Lim, H. J.; Yang, S. J.; Park, C. R. Surface modifications for the effective dispersion of carbon nanotubes in solvents and polymers. Carbon 2012, 50, 3−33.

(20) Skender, A.; Hadj-Ziane-Zafour, A.; Flahaut, E. Chemical functionalization of Xanthan gum for the dispersion of double-walled carbon nanotubes in water. Carbon 2013, 62, 149−156.

(21) Vaisman, L.; Wagner, H. D.; Marom, G. The role of surfactants in dispersion of carbon nanotubes. Adv. Colloid Interface Sci. 2006, 128, 37−46.

(22) Yu, J.; Grossiord, N.; Koning, C. E.; Loos, J. Controlling the dispersion of multi-wall carbon nanotubes in aqueous surfactant solution. Carbon 2007, 45, 618−623.

(23) Blanch, A. J.; Lenehan, C. E.; Quinton, J. S. Optimizing surfactant concentrations for dispersion of single-walled carbon nanotubes in aqueous solution. J. Phys. Chem. B 2010, 114, 9805−9811.

(24) Wang, H.; Zhou, W.; Ho, D.; Winey, K.; Fischer, J.; Glinka, C.; Hobbie, E. Dispersing single-walled carbon nanotubes with surfactants: A small angle neutron scattering study. Nano Lett. 2004, 4, 1789−1793.

(25) Vigolo, B.; Penicaud, A.; Coulon, C.; Sauder, C.; Pailler, R.; Journet, C.; Bernier, P.; Poulin, P. Macroscopic fibers and ribbons of oriented carbon nanotubes. Science 2000, 290, 1331−1334.

(26) Jiang, L.; Gao, L.; Sun, J. Production of aqueous colloidal dispersions of carbon nanotubes. J. Colloid Interface Sci. 2003, 260, 89−94.

(27) Shen, Z.; Wang, G. Colloids and Surfaces Chemistry; Chemical Industry Press: Peking, 1991.

(28) Chattopadhyay, D.; Galeska, L.; Papadimitrakopoulos, F. A route for bulk separation of semiconducting from metallic single-wall carbon nanotubes. J. Am. Chem. Soc. 2003, 125, 3370−3375.

(29) Lewenstein, J.; Burgin, T.; Ribayrol, A.; Nagahara, L.; Tsui, R. High-yield selective placement of carbon nanotubes on pre-patterned electrodes. Nano Lett. 2002, 2, 443−446.

(30) Chen, S.; Jiang, Y.; Wang, Z.; Zhang, X.; Dai, L.; Smet, M. Light-controlled single-walled carbon nanotube dispersions in aqueous solution. Langmuir 2008, 24, 9233−9236.

(31) Ding, Y.; Chen, S.; Xu, H.; Wang, Z.; Zhang, X.; Ngo, T. H.; Smet, M. Reversible dispersion of single-walled carbon nanotubes based on a $CO_2$-responsive dispersant. Langmuir 2010, 26, 16667−16671.

(32) Zhong, W.; Claverie, J. P. Probing the carbon nanotube-surfactant interaction for the preparation of composites. Carbon 2013, 51, 72−84.

(33) Camponeschi, E.; Florkowski, B.; Vance, R.; Garrett, G.; Garmestani, H.; Tannenbaum, R. Uniform directional alignment of single-walled carbon nanotubes in. viscous polymer flow. Langmuir 2006, 22, 1858−1862.

(34) Matarredona, O.; Rhoads, H.; Li, Z.; Harwell, J.; Balzano, L.; Resasco, D. Dispersion of single-walled carbon nanotubes in aqueous solutions of the anionic surfactant NaDDBS. J. Phys. Chem. B 2003, 107, 13357−13367.

(35) Fernando, K.; Lin, Y.; Wang, W.; Kumar, S.; Zhou, B.; Xie, S.; Cureton, L.; Sun, Y. Diminished band-gap transitions of single-walled carbon nanotubes in complexation with aromatic molecules. J. Am. Chem. Soc. 2004, 126, 10234−10235.

(36) Zhang, J.; Lee, J.; Wu, Y.; Murray, R. Photoluminescence and electronic interaction of anthracene derivatives adsorbed on sidewalls of single-walled carbon nanotubes. Nano Lett. 2003, 3, 403−407.

(37) Moore, V.; Strano, M.; Haroz, E.; Hauge, R.; Smalley, R.; Schmidt, J.; Talmon, Y. Individually suspended single-walled carbon nanotubes in various surfactants. Nano Lett. 2003, 3, 1379−1382.

(38) Xin, X.; Xu, G.; Zhang, Z.; Chen, Y.; Wang, F. Aggregation behavior of star-like PEO-PPO-PEO block copolymer in aqueous solution. Eur. Polym. J. 2007, 43, 3106−3111.

(39) Zhang, Z.; Xu, G.; Fang, W.; Dong, S.; Chen, Y. Demulsification by amphiphilic dendrimer copolymers. J. Colloid Interface Sci. 2005, 282, 1−4.

(40) Dror, Y.; Pyckhout-Hintzen, W.; Cohen, Y. Conformation of polymers dispersing single-walled carbon nanotubes in water: A small-angle neutron scattering study. Macromolecules 2005, 38, 7828−7836.

(41) Sun, J.-T.; Hong, C.-Y.; Pan, C.-Y. Surface modification of carbon nanotubes with dendrimers or hyperbranched polymers. Polym. Chem. 2011, 2, 998−1007.

(42) Vaisman, L.; Marom, G.; Wagner, H. Dispersions of surface-modified carbon nanotubes in water-soluble and water-insoluble polymers. Adv. Funct. Mater. 2006, 16, 357−363.

(43) Islam, M.; Rojas, E.; Bergey, D.; Johnson, A.; Yodh, A. High weight fraction surfactant solubilization of single-wall carbon nano-tubes in water. Nano Lett. 2003, 3, 269−273.

(44) Strano, M.; Moore, V.; Miller, M.; Allen, M.; Haroz, E.; Kittrell, C.; Hauge, R.; Smalley, R. The role of surfactant adsorption during ultrasonication in the dispersion of single-walled carbon nanotubes. J. Nanosci. Nanotechnol. 2003, 3, 81−86.

(45) Jayamurugan, G.; Vasu, K. S.; Rajesh, Y. B. R. D.; Kumar, S.; Vasumathi, V.; Maiti, P. K.; Sood, A. K.; Jayaraman, N. Interaction of single-walled carbon nanotubes with poly(propyl ether imine) dendrimers. J. Chem. Phys. 2011, 134.

(46) Sunder, A.; Hanselmann, R.; Frey, H.; Mulhaupt, R. Controlled synthesis of hyperbranched polyglycerols by ring-opening multi-branching polymerization. Macromolecules 1999, 32, 4240−4246.

(47) Catalog Sigma-Aldrich, 2014; http://www.sigmaaldrich.com/ cataloge/.

(48) Grossiord, N.; van der Schoot, P.; Meuldijk, J.; Koning, C. E. Determination of the surface coverage of exfoliated carbon nanotubes by surfactant molecules in aqueous solution. Langmuir 2007, 23, 3646−3653.

(49) Sun, Z.; Nicolosi, V.; Rickard, D.; Bergin, S. D.; Aherne, D.; Coleman, J. N. Quantitative evaluation of surfactant-stabilized single-walled carbon nanotubes: Dispersion quality and its correlation with zeta potential. J. Phys. Chem. C 2008, 112, 10692−10699.

(50) Rastogi, R.; Kaushal, R.; Tripathi, S. K.; Sharma, A. L.; Kaur, I.; Bharadwaj, L. M. Comparative study of carbon nanotube dispersion using surfactants. J. Colloid Interface Sci. 2008, 328, 421−428.

(51) Garamus, V.; Maksimova, T.; Kautz, H.; Barriau, E.; Frey, H.; Schlotterbeck, U.; Mecking, S.; Richtering, W. Hyperbranched polymers: Structure of hyperbranched polyglycerol and amphiphilic

poly(glycerol ester)s in dilute aqueous and nonaqueous solution. Macromolecules 2004, 37, 8394−8399.

(52) Haggenmueller, R.; Rahatekar, S. S.; Fagan, J. A.; Chun, J.; Becker, M. L.; Naik, R. R.; Krauss, T.; Carlson, L.; Kadla, J. F.; Trulove, P. C.; Fox, D. F.; DeLong, H. C.; Fang, Z.; Kelley, S. O.; Gilman, J. W. Comparison of the quality of aqueous dispersions of single wall carbon nanotubes using surfactants and biomolecules. Langmuir 2008, 24, 5070−5078.

(53) Napper, D. Polymeric Stabilization of Colloidal Dispersion; Academic Press: London, 1983.

(54) Callens, M.; De Greef, N.; Seo, J.; Gorbatikh, L.; Verpoest, I. Ductile steel fibre composites with carbon nanotubes grafted on the fibres; Proceedings of the CompositeWeek@Leuven, Leuven, Belgium, 2013.

(55) Gong, X.; Liu, J.; Baskaran, S.; Voise, R.; Young, J. Surfactant-assisted processing of carbon nanotube/polymer composites. Chem. Mater. 2000, 12, 1049−1052.